# Surface transport and band gap structure of exfoliated 2H-$MoTe_2$ crystals

Ignacio Gutiérrez Lezama[1], Alberto Ubaldini[2], Maria Longobardi[2], Enrico Giannini[2], Christoph Renner[2], Alexey B. Kuzmenko[2] and Alberto F. Morpurgo[1].

1 DPMC and GAP, Université de Genéve, 24 quai Ernest Ansermet, CH-1211 Geneva, Switzerland.
2 DPMC, Université de Genéve, 24 quai Ernest Ansermet, CH-1211 Geneva, Switzerland.

**Abstract**
Semiconducting transition metal dichalcogenides (TMDs) have emerged as materials that can be used to realize two-dimensional (2D) crystals possessing rather unique transport and optical properties. Most research has so far focused on sulfur and selenium compounds, while tellurium-based materials attracted little attention so far. As a first step in the investigation of Te-based semiconducting TMDs in this context, we have studied $MoTe_2$ crystals with thicknesses above 4 nm, focusing on surface transport and a quantitative determination of the gap structure. Using ionic-liquid gated transistors, we show that ambipolar transport at the surface of the material is reproducibly achieved, with hole and electron mobility values between 10 and 30 $cm^2/Vs$ at room temperature. The gap structure is determined through three different techniques: ionic-liquid gated transistors and scanning tunneling spectroscopy, that allow the measurement of the indirect gap ($E_{ind}$), and optical transmission spectroscopy on crystals of different thickness, that enables the determination of both the direct ($E_{dir}$) and the indirect gap. We find that at room temperature $E_{\mathrm{ind}}$ = 0.88 eV and $E_{dir}$ = 1.02 eV. Our results suggest that thin $MoTe_2$ layers may exhibit a transition to a direct gap before mono-layer thickness. They should also drastically extend the range of direct gaps accessible in 2D semiconducting TMDs.

## Introduction

Semiconducting transition metal dichalcogenides (TMDs) [1] are attracting interest in the field of opto-electronics [2]–[6], for several reasons. Many of them quite generically support electron and hole conduction with balanced carrier mobility [6]–[9], a feature essential in the realization of devices for light generation and light conversion. Their semiconducting band gap can be tuned [10]–[16] over a broad range, by either choosing one of the numerous existing compounds [13], [14], [17]–[19] (semiconducting TMDs have chemical formula $MX_2$, where M is a transition metal and X a chalcogen; in the best known ones M = Mo or W and X = S, Se or Te), or through size quantization [10]–[12], [16]. Indeed, upon reducing the thickness of the material, the indirect band-gap present in bulk compounds increases by 0.5 - 0.7 eV depending on the material [11], [12], becoming –for monolayers- larger than the direct gap [11], [12], [16]. Additionally, the van der Waals bonds that hold the $MX_2$ layers together facilitate the combination of different TMDs into heterostructures [2], [14], [20], offering considerable potential for the realization of opto-electronic devices [2], [20].

Among all semiconducting TMDs, Te-based crystals of nanometer thickness have remained virtually unexplored. In particular, this is the case for $MoTe_2$. Although the electronic properties of this material were studied in the past [21]–[25], the focus of earlier investigations has been on bulk transport [23], [24] and thermoelectric phenomena [21], [22]. With the exception of some angle-resolved photoemission experiments [25] and optical transmission measurements through thick crystals [21], no systematic and reliable spectroscopy has been performed. Additionally, to our knowledge, no work on nano-structured devices –of the type of interest in opto-electronics- or studies based on few-layer thick flakes have been reported. Experiments are therefore needed both to assess the quality of the surface of $MoTe_2$, crucial in devices based on layers of atomic thickness, and to determine carefully the parameters that govern the basic optical processes and the opto-electronic functionality of the material. As a first step in this direction, we report a systematic study of exfoliated $MoTe_2$ with thickness down to approximately 4 nm.

In particular, we discuss the realization of devices based on crystals having thickness down to a few nanometers, we characterize the surface of these crystals by scanning tunneling microscopy, and perform measurements using ionic-liquid gated transistors to demonstrate that the surface supports ambipolar transport. We then proceed to study the band gap structure of $MoTe_2$, using different techniques. From the transfer curves of the ionic-liquid gated transistors and from scanning tunneling spectroscopy we extract the value of the smallest distance –in energy– between the valence and the conductance bands. The values found with the two techniques are consistent with the result of optical transmission measurements through crystals of different thickness, from which we can independently determine both the direct and the indirect gaps of $MoTe_2$. We find that at room temperature the indirect gap is approximately 0.88 eV and the direct gap is 1.02 eV, so that the difference between the two is approximately 0.15 eV. The implications of this small difference between direct and indirect gaps – four-to-five times smaller than for semiconducting TMDs based on S and Se chalcogen atoms – will be discussed in some detail at the end of the paper.

**Methods**

The $MoTe_2$ single crystals used in this work were grown following a newly developed chemical vapor transport method that uses $MoCl_5$ as a precursor (see reference [26] for details). This method yields 2H-$MoTe_2$ crystals (also known as α-$MoTe_2$) with lateral dimensions of several millimeters, which enabled us to perform different kinds of measurements on flakes extracted from a same $MoTe_2$ parent crystal. The quality of the crystals was evaluated via X-ray diffraction (see ref. [26]) and by Raman spectroscopy (see figure 1(a); the Raman spectrum is in good agreement with previous reports [27]). Both thin (4 – 130 nm) and thick (up to tens of microns) exfoliated flakes (see figure 1(b)) were used. We evaluated the quality of the surface of the $MoTe_2$ flakes via scanning tunneling microscope (STM) topography images. A highly ordered lattice with clear atomic resolution and a low density of defects is observed (figure 1(c)). The hexagonal symmetry observed in both real and reciprocal space (see inset of figure 1(c)) likely originates from the Te atoms, since the Te-Mo-Te layers are cleaved along the van der Waals Te-Te bonds.

Ionic-liquid (IL) gated $MoTe_2$ field-effect transistors (FETs) were fabricated using thin (20 – 40 nm) exfoliated flakes transferred onto $SiO_2$/Si substrates. The optical contrast of the flakes on Si/$SiO_2$ substrates (the $SiO_2$ is 300 nm thick) was calibrated versus the thickness measured using an Asylum Cypher atomic force microscope (AFM) (see figure 1(d)-(f)). Standard electron-beam lithography was employed to create Ti/Au (15/60 nm) contacts in a Hall bar configuration (see figure 2(a)). Two different ionic liquids were used as gate dielectric, either $[P14]^+$ $[FAP]^-$ (1-butyl-1-methylpyrrolidinium tris(pentafluorethyl)trifluorophosphate) or $[DEME]^+$ $[TFSI]^-$ (N,N-Diethyl-N-methyl-N-(2-methoxyethyl)ammonium bis(trifluromethylsulfonyl)imide, with no significant difference in the device operation and quality. An Au gate electrode was used to apply the gate potential, while the potential across the ionic liquid/device interface (the reference potential $V_R$) was measured by introducing an Ag/AgO wire into the ionic liquid (see figure 2(b), and reference [9] for more details). The transport experiments, field-effect and Hall measurements, were performed in the dark, at 270 K and in a He atmosphere.

The STM measurements, both topography and spectroscopy, were carried out in an Omicron LT-STM operating at liquid nitrogen temperature (77 *K*) and an ultrahigh-vacuum (UHV) environment with a base pressure below 1 x $10^{-11}$ mbar. The $MoTe_2$ flakes (μm-thick) were cleaved in vacuum inside the STM chamber, and electrochemically etched W tips, calibrated by measuring the Shockley surface state of a clean Au (111) surface, were used. Finally, the optical transmission spectroscopy measurements (0.5 – 1.3 eV) were performed in ambient conditions using a Bruker Hyperion 2000 infrared microscope coupled to a Bruker Vertex 70v fourier transform spectrometer.

**Results**

As for other semiconducting TMDs that are at the focus of current research [2], the simplest way to produce thin layers of $MoTe_2$ is through exfoliation of bulk crystals using an adhesive tape. Once exfoliated, the thin layers are transferred onto a suitable substrate. To identify the thickness of the flakes it is useful to calibrate their optical contrast $C$ under an optical microscope as a function of thickness ($C = (I_{sub} - I_{flake})/I_{sub}$, where $I_{sub}$ and $I_{flake}$ are the intensity of the substrate and of the flake), as it has been done for other materials [28]–[30]. Figure 1(d) and (e) show an optical microscope image of a flake containing regions of different thickness and the corresponding height profile measured by AFM. The results of similar measurements on many flakes are summarized in figure 1(f), where the contrast is plotted in each of the RGB channels and for white light (inset). When the flake thickness (*t)* is between approximately 4 and 30 nm the contrast provides a pretty accurate and reliable measure. For larger thicknesses the method cannot be used, because the contrast stops changing upon increasing *t*. For $t < 4$ nm (corresponding to approximately 6 mono-layers or less), the height of the step measured by AFM did not a appear to be a reliable measure of the flake thickness because material can be present in between the substrate and the flake (see, for instance, the supplementary information of reference [31]). Additionally, we found that the yield of few-layer thick flakes seems to be lower than the one resulting from the exfoliation of other TMDs. Indeed, we found few-layer thick flakes with sufficiently large lateral dimensions for device fabrication (see inset of figure 1(e)) only towards the end of this work.

We employed exfoliated $MoTe_2$ flakes of different thickness, both to investigate surface transport using ionic-liquid gated field-effect transistors (FETs) [8], [9], and to perform optical transmission spectroscopy (in which case glass was used as substrate). Figure 2(a) shows an optical microscope image of a thin $MoTe_2$ crystal on $Si/SiO_2$, contacted with metallic (Au/Ti) leads in a multi-terminal configuration, to be used for the realization of an ionic-liquid gated FET (see figure 2(b) for the device schematics). We measured transport on approximately 10 different such FETs, and the data shown in figure 2(c)-(g) are representative of their behavior. Clear ambipolar transport was observed in all the devices, both in the transfer curves (i.e., the source-drain current $I_{SD}$ measured versus gate voltage $V_G$, at fixed source-drain bias $V_{SD}$; see figure 2(c)) and in the output curves (i.e., $I_{SD}$ vs $V_{SD}$ at fixed $V_G$; figure 2(f) and (g)). In the latter, ambipolar transport is responsible for the increase in current –rather than the customary saturation- at large values of $V_{SD}$ (see figure 2(f) for $V_G'$= -0.4 V and -0.7 V, and figure 2(g) for $V_G'$=0, 0.2, and 0.4 V). The observation of balanced ambipolar transport implies that the potential barriers formed at the metal/contact interface are sufficiently small (or sufficiently thin due to electrostatic screening of the ions in the ionic-liquid [9]) to provide good injection of both types of charge carriers. This is important for the determination of the band gap, and –as we discuss below in more detail- the value of the gap that we extract from the analysis of the FET characteristics is not affected by the contact resistance.

It is apparent in figure 2(c) that a significant hysteresis is present in the measurements upon sweeping $V_G$ (the current is plotted versus the potential measured at the reference electrode, $V_R$, to account for a possible voltage drop at the gate/ionic liquid interface [9]). Such hysteresis is not uncommon in ionic-liquid gated devices reported in the literature [8] (at least when measurements in both $V_G$ sweep directions are actually shown). In $MoTe_2$ devices, the effect is larger than in analogous devices realized in our laboratory on other semiconducting dichalcogenides: $WS_2$ exhibits normally negligible hysteresis [9], and in $MoS_2$ [8] the hysteresis is larger than in $WS_2$, but still smaller than in $MoTe_2$. We have checked that in the explored gate voltage range the hysteresis is not due to irreversible chemical changes in the material, since no overall device degradation is observed upon extensive sweeping of the gate voltage. Although a definite conclusion about its precise microscopic origin cannot be drawn at this stage, the sensitivity of the hysteresis to the specific materials suggests that it is at least in part related to the material quality. Indeed, among all semiconducting dichalcogenides that we are studying, our $MoTe_2$ crystals appear to have the largest density of unintentional dopants and in-gap states. This can be inferred from the bulk conductivity (i.e., conductivity measured at $V_G$ = 0 V, which in our devices can be of both p- and n-type nature), and from the measured sub-threshold swing $S$ for electron and holes (see figure 2(d)-(e); respectively $S \sim 140$ *mV/dec* and ~125 *mV/dec*). We have also verified that the same values of S, within 5%, are obtained from the four-terminal conductivity, which implies that the contact resistance does not significantly influences the sub-threshold device characteristics (and in particular, it does not determine the value of *S*). Even though the *S*-values are reasonably good in absolute terms for transistors realized on crystals grown in an un-optimized way, they deviate significantly from the ultimate limit $S = kT/e \ln(10)$ [32] that we regularly observe in $WS_2$ [6], [9]. All these observations are consistent with the presence of disorder which generates tails of localized states inside the band-gap, at the bottom of the conduction band and at the top of the valence band. To make a rough estimate of the width $\Delta W$ of these tails (i.e., the extension –in energy- of the region where localized states are present inside the band-gap), we assume that the associated broadening in the density of states at the band edges is responsible for the larger value of *S* in a way similar to the thermal broadening in the electronic population, i.e. $S = \sqrt{(kT/e)^2 + (\Delta W/e)^2}\ln(10)$. Following such a criterion we obtain $\Delta W \approx 50$ meV.

As we demonstrated recently on $WS_2$ [9], the extremely large capacitance of the ionic-liquid gate enables the band gap of the material to be extracted directly from the transfer curve of a FET, measured as a function of reference voltage. To this end, it suffices to measure the difference in threshold voltage for electron and hole transport, and to calculate $\Delta_{ILG} = e\left|V_{T,e} - V_{T,h}\right|$ (the $V_T$'s are obtained by extrapolating the linear part of the transfer curve to $I_{DS} = 0$ *A* [32]). Because of the significantly larger hysteresis, it is not a priori clear whether the same strategy can give a reliable value for the band gap of $MoTe_2$. To check, we estimate $\Delta_{ILG}$ by taking the difference in threshold voltage extracted when sweeping $V_G$ both from positive to negative values, and from negative to positive. We find that both values coincide within a few tenths of meV. We conclude that ionic-liquid gating provides a first estimate of the gap $\Delta_{ILG} = 0.83 \pm 0.05$ eV (as error we take the estimated band tail width). We have also verified that the value of the band gap obtained by applying the same procedure to the four-terminal conductivity is consistent with the value extracted from the two-terminal measurements shown here, which implies that the contact resistance has no effect on the determination of the band gap. Despite the overall consistency of our findings, the reliability and accuracy of the extracted gap value will obviously have to be checked by measuring the gap using more established techniques, such as scanning tunneling microscopy and optical transmission spectroscopy.

Before discussing the results obtained with these techniques, we conclude the part on surface transport, by discussing the measurement of the carrier mobility. To this end, we extracted the carrier density $n_{e,h}$ from Hall effect measurements for different values of $V_G$, corresponding to both electron and hole accumulation. The electron/hole mobility $\mu_{e,h}$ was then calculated from the longitudinal conductivity $\sigma_{e,h} = e\mu_{e,h}n_{e,h}$. The results are shown in figure 2(h). The maximum hole and electron mobility were 10 $cm^2$/Vs and 30 $cm^2$/Vs. These values are comparable to –but somewhat lower than- the field-effect mobility values observed in ionic-liquid gated FETs based on other TMDs [8], [9], confirming that the $MoTe_2$ crystals used in the experiments appear to be more disordered than other semiconducting TMDs crystals.

An obvious experimental technique to measure the band gap of a semiconductor is scanning tunneling spectroscopy (STS). We performed systematic STS measurements at 77 K on different $MoTe_2$ crystals cleaved in UHV. Figure 3 shows differential conductance spectra obtained by numerically differentiating $I(V)$ curves measured at different positions on the crystal surface. The data show very good reproducibility, with a clear gap in the density of states. The reproducibility is indicative of the overall uniformity of the material, despite the presence of defects that appear to be charge centers below the sample surface (not shown), and which seem to shift lightly rather rigidly the density of states in energy.

As in the case of IL gating, STS probes the smallest distance –in energy- between valence and conduction bands, irrespective of whether this gap is direct or indirect. For a quantitative estimate of the band gap from the STS data, we computed an averaged *dI/dV* vs *V* curve, corresponding to the red dashed line in figure 3. Both for negative and positive bias –corresponding to injecting holes in the valence band and electrons in the conduction band, respectively- a linear increase in *dI/dV* as a function of *V* is seen. To define the onset of the band, we extrapolated the linear part of the curve to its crossing with the minimum conductance (see inset of figure 3) [33]. The difference of the two onsets (multiplied by the electron charge) gives the value of the gap. We find $E_{STS} = 0.96 \pm 0.02\ eV$ (the error corresponds to the sample-to-sample fluctuations in the value of the extracted gap). The difference between this value and that found from the analysis of IL FETs –although not large- appears to be significant, i.e. it is somewhat larger than the indetermination associated to the experimental procedures. This difference in gap values likely originates from the lower temperature at which the STS measurements are performed (77 K, as compared to 270 K for the FET experiments), since the gap of TMDs are known to have a weak dependence on T [34]. Indeed, the only previous optical spectroscopy measurements of the indirect gap on thick $MoTe_2$ crystals at the corresponding temperatures [21] agree well with both the STS and the FET measurements (see also discussion below). We conclude that the ionic-liquid gated FET devices and the STS measurements provide consistent values for the smallest gap in $MoTe_2$.

In the attempt to measure also the direct band gap, we have performed optical transmission spectroscopy measurements. Several crystals of varying thickness were used, which is essential for different reasons. In very thick crystals (> 1 μm) absorption due to the indirect transition completely suppresses the amplitude of the transmitted signal, preventing the observation of the optical transition associated to the direct gap that occurs at higher energy. In thinner crystals the situation is opposite: the change in optical absorption due to the direct gap transition becomes measurable, but the signal originating from indirect (phonon-mediated) transitions is too small to measure. Additionally, a quantitative determination of the direct gap requires a careful fitting of the frequency dependent transmission to extract the dielectric function. In practice, the procedure is complicated due to the Fabry-Perot interference caused by the finite thickness of the $MoTe_2$ crystal and the presence of the substrate. To ensure the correctness of the final result, therefore, it is important to show that the same Kramers-Kroning consistent dielectric function reproduces well the data measured on crystals of different thickness using an appropriate multilayer optical model.

We have measured optical transmission through layers exfoliated form a same parent $MoTe_2$ crystal, having thickness of 20 μm, 120 nm, and 32 nm, as determined by AFM. The frequency dependent transmission data are shown in figure 4(a). The thicker 20 μm crystal shows absorption starting at the frequency that corresponds well to the indirect gap, as expected, and allows us to fix unambiguously the contribution of the indirect transition to the dielectric function of $MoTe_2$ (for this crystal thickness the period of the Fabry-Perot oscillations below the band gap is very short and can be easily removed before starting the analysis; indeed, these oscillations have been removed from the spectrum shown in figure 4(a)). As expected, in the thicker crystal the absorption at frequencies higher than the indirect gap transition is too strong to allow the observation of the optical transition associated to the direct gap. On the contrary, for the $t$ = 120 and 30 nm crystals, the transmitted signal can be easily detected up to much higher frequencies, enabling the determination of the dielectric constant.

The complex dielectric function ε ($h\nu$) of $MoTe_2$ (figure 4(b)) was obtained using a simultaneous Kramers-Kroning constrained variational fitting [35] of the transmission spectra of the three crystals. The spectra were initially fitted using a limited number of Drude-Lorentz and Tauc-Lorentz functions to model the overall shape of the transmission curves. After that, a variational dielectric function, which consists of a large number of narrow oscillators at fixed evenly spaced frequencies and with variable intensities, was added on top and adjusted in order to reproduce the detailed spectral structure of the three transmission curves.

From the dielectric function we obtain the optical absorption coefficient $\alpha$ ($h\nu$), shown in figure 4(c). Significant absorption starts from approximately 0.9 $eV$ (see figure 4(c) and its inset), and increases pronouncedly above approximately 1 $eV$. The data are consistent with the functional dependence expected for indirect and direct gaps. Specifically, for an indirect gap $\alpha(h\nu) \propto (h\nu - E_g^i \pm E_{ph})^2$, ($E_g^i$ is the indirect gap and $E_{ph}$ is the energy of the phonon that ensures momentum conservation in the indirect transition) so that just after the absorption onset $\alpha^{1/2}$ is expected to scale linearly with energy. For a direct gap, $\alpha(h\nu) \propto \left(h\nu - E_g^d\right)^{1/2}$ , so that a linear scaling is expected when plotting $\alpha^2$ ($E_g^d$ is the direct gap; these functional dependencies of $\alpha(h\nu)$ hold for 3D semiconductors with a parabolic dispersion relation, even in the in the presence of strong anisotropy, which is the case for the $MoTe_2$ flakes of the thicknesses investigated here. For truly 2D electronic systems –such as $MoTe_2$ monolayers- the functional form of $\alpha(h\nu)$ would be different). Figure 4(d) and (e) show that the data indeed exhibit the expected scaling for both transitions, enabling their identification. By extrapolating the plots to zero absorption we determine that $E_g^i \pm E_{ph} = 0.88\ eV$, and that $E_g^d = 1.02\ eV$. From this analysis we can conclude that an indirect gap $E_g^i = 0.88\ \pm 0.05\ eV$ (where the error is mainly associated to the estimated, unknown energy of the phonon involved in the indirect transition) and a direct gap $E_g^d = 1.02\ \pm 0.05\ eV$ do coexist in bulk $MoTe_2$.

We conclude that the three different techniques that we have used –transport through ionic-liquid gated FETs, scanning tunneling spectroscopy and optical transmission spectroscopy- give consistent results for the structure of the gap of $MoTe_2$. At room temperature this material has an indirect gap of 0.88 eV and a direct one of 1.02 eV. At least a few aspects of the consistency of these results are worth noting. One is the reliability of the gap value measured with ionic liquid FETs, which –as we have shown here- has given the correct result despite the fairly significant device non-ideality. This observation is remarkable, since the use of ionic-liquid gated FETs to extract quantitatively the gap of semiconductors has been proposed only recently [9] and it is important to test how far the values extracted with this technique are reliable. The second is the fact that quantitative measurements of the gap are often not straightforward, because different effects can play a role. For instance, in optical measurements, excitonic effects can make the identification of the onset of inter-band optical transition difficult. Similarly, in scanning tunneling spectroscopy, the effect of the bias voltage applied to the tip can induce uncontrolled band bending at the material surface (see for instance [36]), affecting the onset of the tunneling current, and thereby the extracted gap value. Our results show that within an experimental error of approximately 5% (corresponding to an uncertainty of approximately 50 meV) all techniques give fully consistent results.

**Discussion and Conclusions**

The results discussed above indicate that $MoTe_2$ can play an important role in extending the potential of semiconducting TMD materials for opto-electronic applications. The transport measurements confirm that –in line with results obtained on other TMD semiconductors- $MoTe_2$ supports gate-induced ambipolar transport at its surface. For the carrier density values explored in our experiments, simple electrostatic screening arguments allow to estimate the depth of the region in which carriers are accumulated to approximately 1 nanometer, corresponding to a couple of $MoTe_2$ layers at the most. It should therefore be expected that ambipolar transport will be present in $MoTe_2$ even when the material is exfoliated down to atomic scale thickness, as observed in other semiconducting TMDs. The carrier mobility values that we observed are not drastically smaller than those of other semiconducting TMDs, despite the fact that the crystal growth has not been optimized to enhance the semiconducting quality. It therefore appears that $MoTe_2$ is suitable for the realization of ambipolar structures, forming the basis for devices such as light-emitting diodes, photo-detectors, and solar cells.

This finding is particularly interesting in view of the possibility to access $MoTe_2$ crystals only one or few monolayers thick. Indeed, it has been found that in all the semiconducting TMDs investigated so far the bulk indirect band gap is significantly smaller than the direct one [17], and that upon thinning down the material the indirect gap increases (by as much as 0.7 eV for $MoS_2$ [11] and for $WS_2$ [6]), while the direct gap remains approximately constant. As a consequence, all these different TMDs exhibit a direct band gap close to 2 eV (the precise value depends on the material), when reduced to individual monolayers [12]. We find that bulk $MoTe_2$ has qualitatively the same gap structure as the other TMDs –an indirect gap smaller than the direct one- but that quantitatively the situation is different, for two important reasons. The first is that the difference between indirect and direct bulk gap is only approximately 0.15 eV, i.e., four-to-five times smaller than in the semiconducting TMDs that have been studied in monolayer form [11]. The second is that the direct gap is relatively small, approximately 1 eV, much smaller than in the other semiconducting TMDs [17].

These conclusions have some interesting implications, especially if the scaling of the direct and indirect gaps of $MoTe_2$ with thinning down the material follows the same trend observed in all other semiconducting TMDs. Because of the small difference in size between the bulk direct and indirect gaps, a direct gap may already be obtained at thicknesses larger than one monolayer, making $MoTe_2$ different from the other semiconducting TMDs investigated so far. Additionally, since in the other TMDs the direct gap is not varying much with thickness and remains comparable to the bulk value, it should be expected that $MoTe_2$ crystals of atomic thickness would have a direct gap close to 1 eV. Being approximately 2 times smaller than in other semiconducting TMD monolayers, this would broaden considerably the range of gap values accessible in this class of materials, resulting in a drastic extension of the frequency range useful for optical processes. Such a broader range of band gaps would be relevant for technology, and could be particularly interesting in the development of TMD heterostructures [2], [4], [20]. Because of all these considerations, our results therefore indicate that extending the investigation of $MoTe_2$ –and more in general of other Te-based TMDs- to atomically thin layers is a direction certainly worth pursuing.

**Acknowledgments**

We gratefully acknowledge A. Ferreira for technical assistance and N. Ubrig for providing the Raman spectroscopy data. Financial support from the Swiss National Science Foundation (this work is part of the SNF Sinergia project "Electronic properties of new atomically thin semiconductors") and from the EU Graphene Flagship is also gratefully acknowledged.

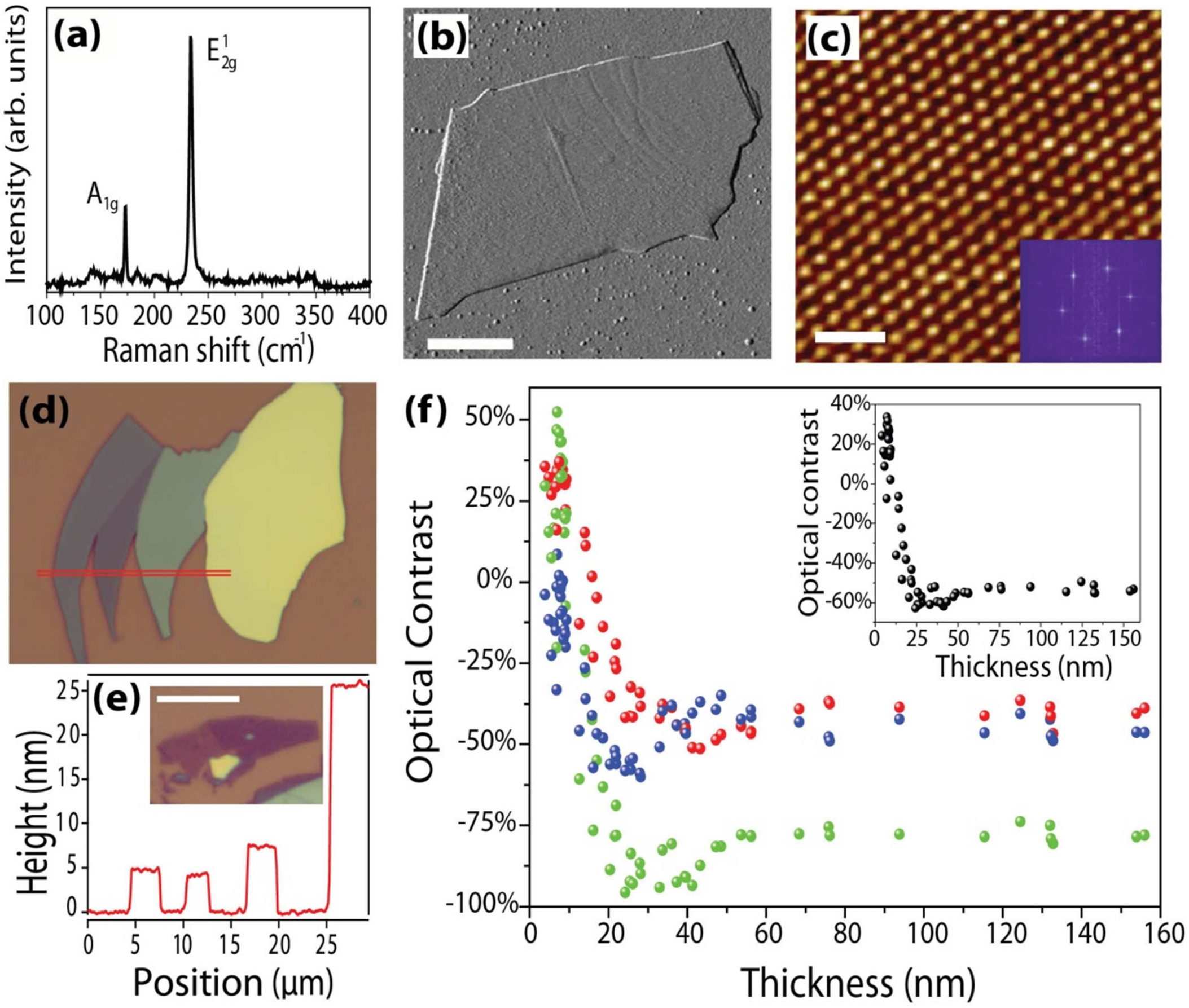

Figure 1. (a) Raman spectrum of a $MoTe_2$ bulk crystal, measured at ambient conditions. (b) Atomic force microscope image of a 20-nm thick exfoliated $MoTe_2$ flake transferred onto a $SiO_2$/Si substrate. The scale bar is 1 μm. (c) Scanning tunneling microscope topography image (sample–tip bias $V_b = -2$ *V*, $I = 50$ *pA*, 77 K) of the surface ($5 \times 5$ nm$^2$ area) of a $MoTe_2$ crystal cleaved in vacuum. The hexagonal symmetry of the surface is clearly observed. The inset shows the unit cell of the reciprocal lattice, as obtained from a Fourier transform of the STM data. The scale bar is 1 nm. (d) Optical microscope image of a $MoTe_2$ flake exhibiting regions of different thickness, and correspondingly different contrast. The AFM height profile shown in (e) was taken along the red line in (d). The inset shows a few-layer thick flake with large lateral dimensions (the scale bar is 10 μm). (f) Optical contrast of $MoTe_2$ flakes as a function of their thickness (measured by AFM), for the red, green and blue intensities (symbols of corresponding color), and for the total intensity (inset).

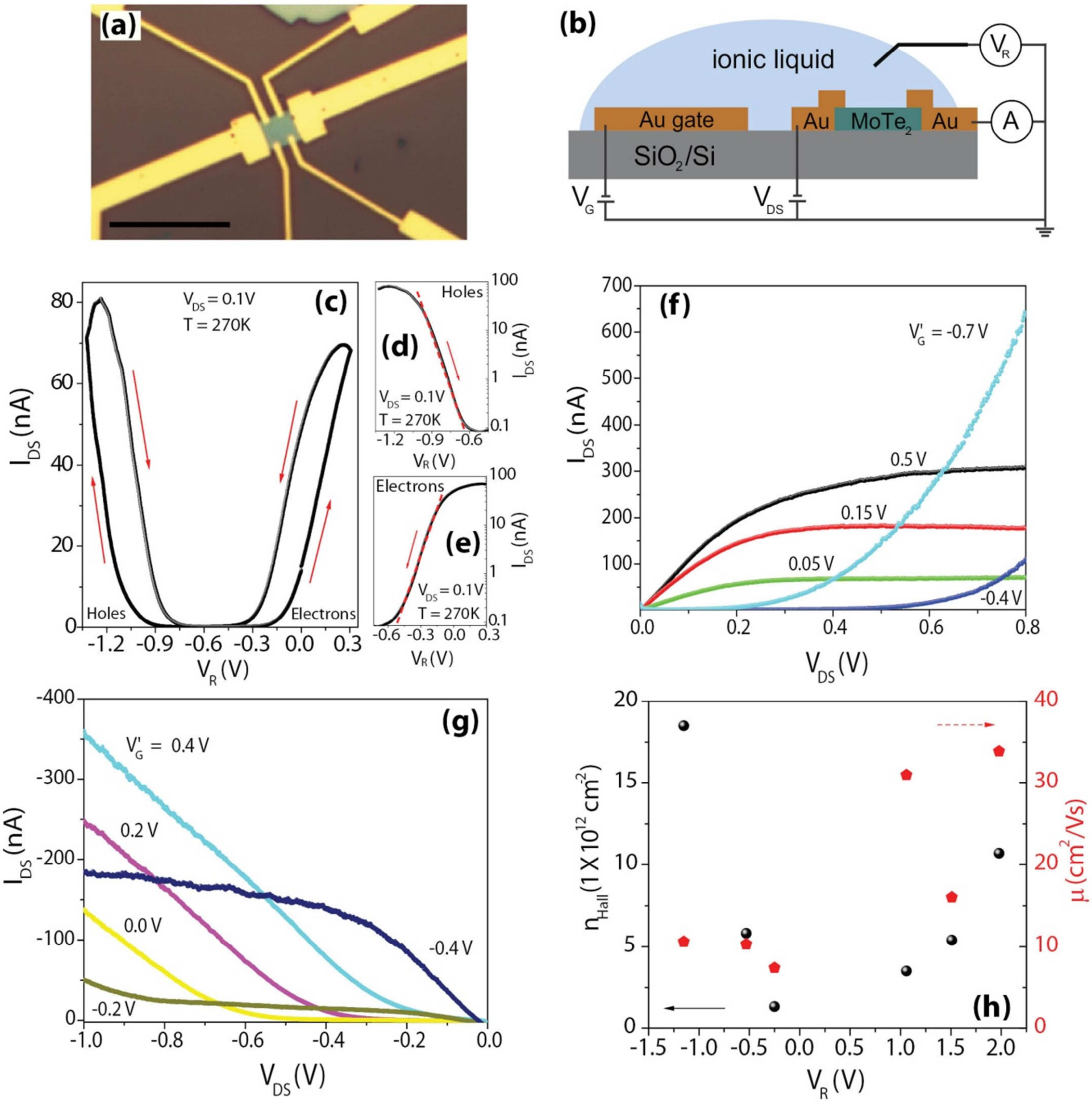

Figure 2. (a) Optical microscope image of a $MoTe_2$ thin flake with Au/Ti contacts patterned in a Hall bar configuration. The scale bar is 5 μm. (b) Schematic view of an ionic-liquid gated $MoTe_2$ FET. (c) Transfer characteristics as a function of reference potential $V_R$, measured at $V_{DS} = 0.1\ V$ and at 270 $K$ (the gate voltage is swept at a 3 mV/s rate; the arrows indicate the sweep direction). (d) and (e) show a log-scale plot of $I_{DS}$ in the sub-threshold region for holes (d) and electrons (e), as it is needed to determine the sub-threshold swing $S$. Only one sweep direction is shown for clarity (the dashed lines are the linear regressions performed to extract $S$). (f) and (g) show the output characteristics of a $MoTe_2$ ionic-liquid gated FET for several values of $V'_G = V_G - V_T$, (in (f) and (g) $V_T$ is the threshold voltage for electrons and holes, respectively). (h) Carrier density (circles) and mobility (stars) at the surface of the $MoTe_2$ flake, as a function of reference potential.

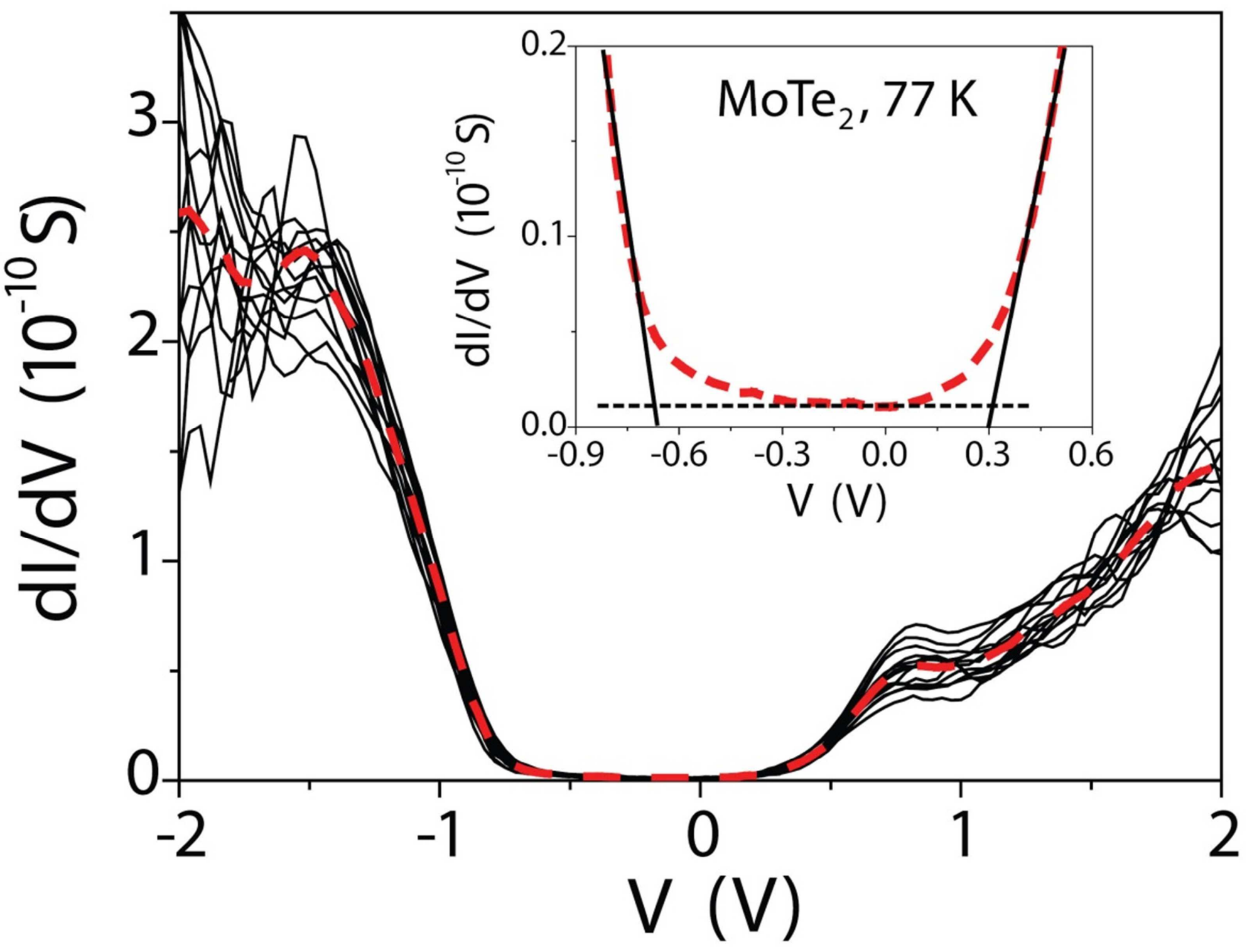

Figure 3. Differential conductance *dI/dV* as a function of sample-tip voltage *V*, measured at 77 K, at different positions on the surface of a $MoTe_2$ flake cleaved in vacuum. The red dashed line is the average of all curves. The inset zooms in on the region where the onset of tunneling conductance in the valence and conduction bands is observed, from which the magnitude of the gap is estimated. The full lines represent the linear extrapolations to minimum conductance, which is indicated by the dashed horizontal line.

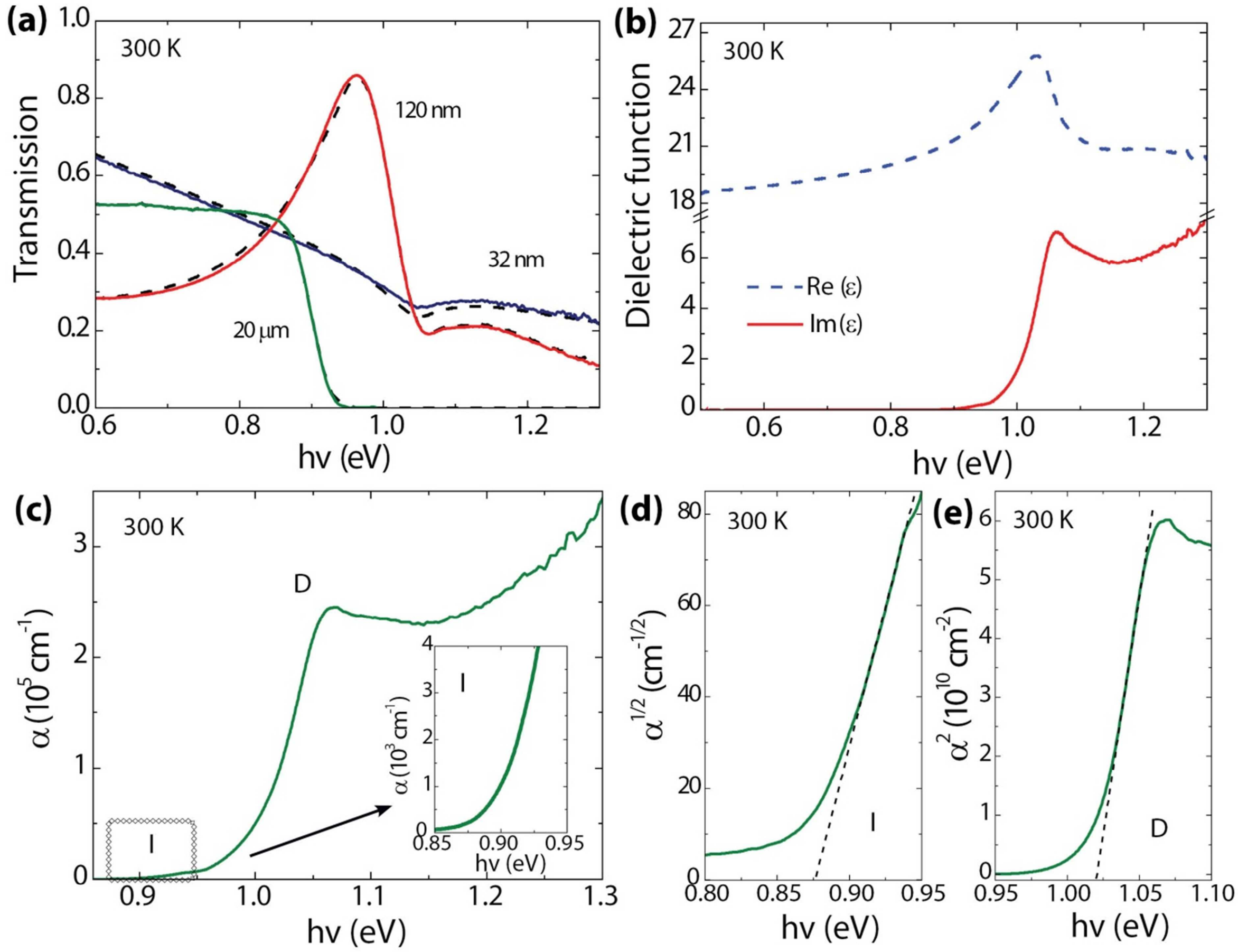

Figure 4. (a) The solid lines represent transmission spectra measured on three exfoliated $MoTe_2$ crystals 32 nm, 120 nm and 20 μm thick, (for the 20 μm thick crystal fast Fabry-Pérot have been removed). The dashed lines are fits performed to obtain the real and imaginary part of the dielectric function shown in (b), from which the absorption coefficient (c) is derived. In (c), I and D mark the features corresponding to absorption due to indirect and direct optical transitions. The inset in (c) shows a magnification of the absorption coefficient at low energy. The dependence of $\alpha^{1/2}$ and $\alpha^2$ on the photon energy ($h\nu$) are shown in (d) and (e)). The dotted lines are the linear extrapolations performed to obtain the band gaps. All data in this figure has been obtained from measurements at 300 K.